\documentclass[twocolumn,tighten]{aastex63_rt42}

\usepackage{multirow}
\usepackage{graphicx}
\usepackage{epstopdf}
\usepackage{amsmath}
\usepackage{psfig}
\usepackage{color}
\usepackage{icomma}

\def\gram{\textrm{g}}

\def\erg{\textrm{erg}}

\def\meter{\textrm{m}}
\def\km{\textrm{km}}

\def\pc{\textrm{pc}}
\def\kpc{\textrm{kpc}}

\def\Myr{\textrm{Myr}}
\def\Gyr{\textrm{Gyr}}

\def\rhoUnits{\textrm{g\ cm}^{-3}}
\def\kms{\textrm{km\ s}^{-1}}

\def\AU{\textrm{au}}

\def\Msun{\textrm{M}_{\odot}}
\def\Mearth{\textrm{M}_{\oplus}}

\def\YJ{\textrm{YJ}}

\def\vrel{v_{\rm rel}}
\def\nbar{\bar{n}}
\def\rbar{\bar{r}}
\def\mbar{\bar{m}}
\def\nISO{n_{\rm ISO}}
\def\nstar{n_{\star}}
\def\rhoStar{\rho_{\star}}
\def\rhoStarSolar{\rho_{\star,\odot}}
\def\GammaImpact{\Gamma_{\rm im}}
\def\Mstar{M_{\star}}
\def\sigmaOne{\sigma_{\rm 1D}}
\def\Kthreat{K_T}
\def\Rprime{R^{\prime}}
\def\adebris{a_{\rm debris}}

\def\bmin{b_{\rm min}}
\def\DeltaThresh{\Delta_{\rm thresh}}
\def\tff{t_{\rm ff}}
\def\te{t_e}
\def\tDelta{t_{\Delta}}

\def\ga{\gtrsim}
\def\la{\lesssim}

\def\endash{\text{--}}

\newcommand{\mean}[1]{\ensuremath{\langle #1 \rangle}}

\def\editbfOne{}
\def\editMathOne{}

\newcommand{\UCB}{Department of Astronomy,  University of California Berkeley, Berkeley CA 94720}

\shorttitle{Elliptical Galaxy Habitability \& Comets}
\shortauthors{Lacki}

\begin{document}

\title{Life in Elliptical Galaxies: Hot Spheroids, Fast Stars, Deadly Comets?}
\correspondingauthor{Brian C. Lacki}
\author[0000-0003-1515-4857]{Brian C. Lacki}
\affiliation{Breakthrough Listen, \UCB}
\email{astrobrianlacki@gmail.com}

\begin{abstract}
Elliptical galaxies have dynamically hot ($\sigmaOne \sim 100 \endash 300\ \kms$) populations of stars, and presumably, smaller objects like comets.  Because interstellar minor bodies are moving much faster, they hit planets harder and more often than in the local Galaxy.  I estimate the rates for Chicxulub-scale impacts on an Earth-size planet in elliptical galaxies as a potential habitability constraint on intelligent life.  Around most stars in a normal elliptical galaxy, these planets receive only $\sim 0.01 \endash 0.1\ \Gyr^{-1}$, although hazardous rates may be common in certain compact early-type galaxies and red nuggets.  About $\sim \editMathOne{5}\%$ of the stellar mass is in a region where the rate is $>10\ \Gyr^{-1}$, large enough to dominate the mass extinction rate.  This suggests that elliptical galaxies have an exclusion zone \editbfOne{of order one hundred} parsecs in radius around their centers for the evolution of intelligent life.
\end{abstract}


\section{Introduction}
\label{sec:Intro}

Galactic habitability is the notion that galactic-scale environmental factors affect the abundance of life-friendly planets.  Thus far, the main identified factors are (1) stellar population metallicity, which can limit the frequency of planets \citep{Gonzalez01,Lineweaver04}; (2) high-energy events (typically associated with massive stars) that may trigger mass extinctions \citep{Clarke81,Annis99,Lineweaver04,Gowanlock16}; and (3) frequent stellar encounters in dense stellar environments \citep{deJuanOvelar12,JimenezTorres13,DiStefano16,Kane18}.

Judging by these criteria, large early-type galaxies (ETGs) should be among the most habitable.  Massive galaxies have high stellar metallicity and likely many planets \citep{Suthar12,Dayal15}.  The lack of recent star-formation reduces the high energy transient threat.  \citet{Whitmire20} speculates their habitability is inhibited by sterilizations early in their history and too high metallicity.  The potential high habitability is especially interesting because they host about half of the stellar mass and likely most terrestrial planets since $z \sim 1$ \citep{Muzzin13,Moffet16,Zackrisson16}.  Their properties are also beneficial for extraterrestrial intelligences (ETIs).  The old stellar populations have had plenty of time for ETI evolution.  Additionally, the hazards for interstellar travel are minimal (\citealt{Lacki21-Traversability}).

Elliptical galaxies are dynamically hot, with stellar velocity dispersions $\sim 100 \endash 300\ \kms$ \citep{Emsellem07}.  Stellar encounters are frequent, but the high relative velocities limit effects of close passages and planetary ejection is unlikely \citep[c.f.,][]{Fregeau06}.  Another hazard comes not from the stars themselves, but what they carry: comets.  If planet formation occurred in elliptical galaxies, there probably is a corresponding population of minor bodies, many ejected as interstellar objects (ISOs)\editbfOne{, as these bodies appear to be a common result of planet formation \citep[e.g.,][]{Hughes18}}.  Impacts from ISOs threaten mass extinction.  The famous Chicxulub impact that triggered the end-Cretaceous extinction released of order $1\ \YJ$ ($10^{31}\ \erg$) of kinetic energy \citep{Alvarez80,Pope97}.\footnote{The high velocity of the impacts may alter the lethality at a given energy, however.}  Of course, even complex life recovers from a mere Chicxulub-like event, with biosphere recovery taking $\sim 1 \endash 30\ \Myr$ depending on taxon and ecological niche \citep{Sahney08,Chen12}. Nonetheless, frequent extinctions are expected to interfere with ETI evolution \citep[c.f.,][]{Cirkovic08}.

Most hazardous impacts on the Earth are from the Solar System's own minor bodies.  The flux rate of these may be modulated by encounters with nearby stars, binary companions, molecular clouds, or Galactic tides \citep{Hills81,Davis84,Duncan87}.  But in elliptical galaxies, the \editbfOne{potential} high fluxes of ISOs add to the threat, compounded because faster objects carry more energy.  Smaller ISOs can trigger mass extinctions, and these are far more common.  A YJ impact could result from a comet with density $0.6\ \rhoUnits$ and radius $\editMathOne{2}\ \km$ hitting at a speed of $\editMathOne{300}\ \kms$.  

This paper considers the habitability constraints for ETIs imposed by ISO impacts in elliptical galaxies. The next section (Section~\ref{sec:ISORate}) derives the impact rate of ISOs with a given kinetic energy on an Earth-sized planet. I present order-of-magnitude estimates of the typical impact rate in Section~\ref{sec:OoM}, and simple models of how the impact rate varies with distance from the galactic center in moderate to large ellipticals in Section~\ref{sec:Models}. \editbfOne{The related threat of orbital perturbation by close stellar passages is considered in Section~\ref{sec:OtherHab}.}

\section{Interstellar comet collision rate}
\label{sec:ISORate}
A biotic world with an impact cross-section $A_p$ is threatened if a minor body hits it with kinetic energy of at least $\Kthreat$.  
The rate of hazardous ISO impacts is found from the relative velocity ($\vrel$) distribution of objects\footnote{At these relative speeds, gravitational focusing is negligible.} and the mass ($m$) distribution density of impactors, assumed to be independent:
\begin{equation}
\GammaImpact (\ge \Kthreat) = \int_0^{\infty} \int_{2\Kthreat/\vrel^2}^{\infty} A_{\editbfOne{p}} \vrel \frac{dP}{d\vrel} \frac{d\nISO}{dm} dm d\vrel .
\end{equation}
The mean space density of objects is the same whether the intruder objects (here all referred to as ISOs) are bound to their host star's Oort cloud or are unbound.

Size distributions of minor bodies are commonly parameterized as a power law in radius, with $dn/dr \propto r^{-q}$ where $q \approx 3 \endash 4.5$.  The mass distribution is then also a power law, $dn/dm \propto m^{-p}$ with $p = (q + 2)/3$.  The discovery of 1I/'Oumuamua and 2I/Borisov have allowed the first empirical determinations of local ISO density.  'Oumuamua has an effective radius \editbfOne{estimated to be less than 130 meters, depending on albedo, shape, and composition \citep[e.g.,][]{Trilling18,Bolin18,Mashchenko19}. I consider an effective radius of $\rbar = 75\ \meter$ \citep{Drahus18} and a comet-like density $\rho = 0.6\ \rhoUnits$, which yields a mass of $(4/3) \pi \rho \rbar^3 = 1.1 \times 10^{12}\ \gram$. T}he estimated density of larger objects \editbfOne{is} $\sim 0.2\ \AU^{-3}$ \citep{Do18}.  Constraints on interstellar meteors and large bodies, including the discovery of 2I/Borisov, \editbfOne{suggest that the size distribution may be even shallower ($q < 4$, $p < 2$), with more large bodies} in this size range \citep{Do18,Jewitt20}. I adopt a power law distribution of $p = 2$ ($q = 4$), with $\nbar = 0.2\ \AU^{-3}$ more massive than $\mbar = \editMathOne{10^{12}}\ \gram$ in the Solar neighborhood.  These are only loose constraints, however, and the actual abundance may differ by an order of magnitude \editbfOne{or more}. 

ISO density presumably scales with stellar population density, with direct proportionality if the mean number of comets formed per star is constant across the Universe.  An unknown $f_c$ variable denotes the effect of other factors on ISO abundance, perhaps the high metallicity in massive galaxies or towards galaxy cores \citep{Henry99}, or the apparent bottom-heavy initial mass function in elliptical galaxies with high stellar velocity dispersions \citep{Ferreras13}.  Quiescent elliptical galaxies are also missing high-mass stars relative to the Milky Way because none have formed recently; any ISOs associated with these past stars likely survive \editbfOne{\citep[as in][]{Veras11,Hansen17}}.  

The ISO mass density distribution is
\begin{equation}
\frac{d\nISO}{dm} = (p - 1) \frac{f_c \rhoStar}{\rhoStarSolar} \frac{\nbar}{\mbar} \left(\frac{m}{\mbar}\right)^{-p} ,
\end{equation}
where $\rhoStarSolar = 0.05\ \Msun\ \pc^{-3}$ is the stellar mass density in the Solar neighborhood \citep{Chabrier01}. \editbfOne{This distribution amounts to $80\,[\log_{10}(m_{\rm max}/m_{\rm min})/6]\ \Mearth$ of ISOs per $\Msun$ of stellar mass in the Solar neighborhood, which as \citet{Do18} notes, is quite high and implies efficient ISO ejection from all stars. The high abundance may be the result of an overestimation of 'Oumuamua's mass, 'Oumuamua belonging to a different population of objects than kilometer-scale comets \citep[e.g.,][]{Jackson21}, the local $\nbar$ being overestimated by orders of magnitude \citep{MoroMartin19-ExoOort}, or efficient production of ISOs around most local stars \citep{Trilling17,Rice19}. The ISO density is well within gross constraints on the metal budget of the Galaxy, however. For the purposes of this paper, $d\nISO/dm$ also includes objects still bound to their host stars, because those too have a relative velocity to other stars $\sim 200\ \kms$. Thus if this is approximately the mass of surviving planetesimals per unit stellar mass in elliptical galaxies, it does not matter if ejection is inefficient. If the planetesimals are typically located in compact debris belts, then the impacts will not occur at a uniform rate, but as rapid series of multiple impacts separated by the time between debris belt passages.}

I model elliptical galaxies as isotropic spheroids with isotropic stellar velocity distributions.  The relative speed distribution is given by a Maxwellian distribution with characteristic velocity $\sqrt{2} \sigmaOne$, where $\sigmaOne$ is the local velocity dispersion along one axis \citep[e.g.,][]{Binney87}:
\begin{equation}
\frac{dP}{d\vrel} = \frac{\vrel^2}{2\sqrt{\pi} \sigmaOne^3} \exp\left(-\frac{\vrel^2}{4 \sigmaOne^2}\right) .
\end{equation}
\editbfOne{The mean $\vrel$ is $(4/\sqrt{\pi}) \sigmaOne$.}

With these distributions, I find that the rate of ISO impacts with kinetic energy $\ge \Kthreat$ is:
\begin{equation}
\GammaImpact (\ge \Kthreat) = \frac{2^{p+1} \Gamma(p + 1)}{\sqrt{\pi}} \frac{f_c \rhoStar}{\rhoStarSolar} \nbar A_{\editbfOne{p}} \left(\frac{\Kthreat}{\mbar}\right)^{1-p} \sigmaOne^{2p - 1} .
\end{equation}
The $\Gamma(p+1)$ term refers to the gamma function.  I find $\GammaImpact \approx 0.1\ \Gyr^{-1}$ if $\sigmaOne = 200\ \kms$ and the ISO number density is equal to that in the Solar neighborhood.  
Given the large uncertainties, I adopt
\begin{equation}
\label{eqn:GammaImpact}
\GammaImpact (\ge \YJ) = \editbfOne{0.017}\ \Gyr^{-1}\ \left(\frac{\rhoStar}{\rhoStarSolar}\right) \left(\frac{\sigmaOne}{200\ \kms}\right)^{\editbfOne{3}}
\end{equation}
as the impact rate, under the assumption that the target worlds are Earth-sized. Since $\GammaImpact$ has a roughly cubic dependence on $\sigmaOne$, the high velocity dispersions of elliptical galaxies increase $\GammaImpact$ by a factor of several hundred relative to the Solar neighborhood.  Larger planets face more frequent impacts, although it is possible that smaller planets are more easily devastated.  

\section{Order-of-magnitude estimate of characteristic impact rate}
\label{sec:OoM}
To get a rough sense of the hazardous ISO impact rate, we can use the central velocity dispersion $\sigma_c$ and mean stellar density $\rho_{1/2}$ within the half-light radius.  \citet{Norris14} has assembled stellar masses, $\sigma_c$, and projected half-light radii $R_e$ for a large variety of stellar systems.  The physical three-dimensional half-light radii $r_{1/2}$ of elliptical galaxies is approximately $(4/3) R_e$ (\citealt{Dehnen93}, D93), so the mean stellar density within $r_{1/2}$ is $\sim (1/2) \Mstar / \editMathOne{[(}4/3\editMathOne{)} \pi r_{1/2}^3\editMathOne{]} \approx 0.05 \Mstar / R_e^3$.  Then from equation~\ref{eqn:GammaImpact}, I estimate a characteristic impact rate of
\begin{equation}
\overline{\GammaImpact} = \editbfOne{0.017}\ \Gyr^{-1}\ \left(\frac{(\Mstar/R_e^3)}{1\ \Msun\,\pc^{-3}}\right) \left(\frac{\sigmaOne}{200\ \kms}\right)^{\editbfOne{3}} .
\end{equation}

\begin{deluxetable*}{lccccc}
\tabletypesize{\footnotesize}
\tablecolumns{4}
\tablewidth{0pt}
\tablecaption{Characteristic hazardous ISO impact rates \label{table:MeanGamma}}
\tablehead{\colhead{Object/class} & \colhead{$\Mstar$} & \colhead{$R_e$} & \colhead{$\overline{\rhoStar}$} & \colhead{$\sigma_c$} & \colhead{$\overline{\GammaImpact}$} \\ & ($10^{10} \Msun$) & ($\kpc$) & \colhead{($\Msun\ \pc^{-3}$)} & \colhead{($\kms$)} & \colhead{($\Gyr^{-1}$)}}
\startdata
Local Milky Way                                       & \nodata & \nodata & $0.05$ & $30$\tablenotemark{\ensuremath{\star}} & \editbfOne{5.7E-5}\\
\hline
E/S0 ($10^9 \le \Mstar \le 10^{10}\ \Msun$)           &  $0.73$ ($0.67 \endash 0.92$)          & $1.8$ ($1.4 \endash 2.3$)           & $0.06$ ($0.03 \endash 0.1$) & $75$ ($67 \endash 86$)    & \editbfOne{2E-3 (4E-4$\endash$ 4E-3)}\\
E/S0 ($10^{10} \le \Mstar \le 10^{11}\ \Msun$)        & $3.4$ ($1.9 \endash 5.8$)             & $2.0$ ($1.4 \endash 2.7$)           & $0.2$ ($0.09 \endash 0.5$)  & $150$ ($110 \endash 180$) & \editbfOne{$0.02$ (8E-3$\endash 0.09$)}\\
E/S0 ($\Mstar \ge 10^{11}\ \Msun$)                    & $24$ ($14 \endash 41$)                & $5.5$ ($3.6 \endash 7.4$)           & $0.08$ ($0.03 \endash 0.1$) & $230$ ($200 \endash 270$) & \editbfOne{$0.04$ ($0.02 \endash 0.07$)}\\
M49                                                   & $60$  & $8.7$  & $0.046$  & $290$ & $\editMathOne{0.047}$\\
M84                                                   & $34$  & $5.5$  & $0.10$   & $290$ & $\editMathOne{0.10}$\\
M87                                                   & $48$  & $6.6$  & $0.085$  & $310$ & $\editMathOne{0.11}$\\
M60                                                   & $45$  & $6.4$  & $0.085$  & $310$ & $\editMathOne{0.11}$\\
NGC 4660                                              & $2.6$ & $0.94$ & $1.6$    & $220$ & $\editMathOne{0.68}$\\
\hline
CSS ($10^6 \le \Mstar \le 10^7\ \Msun$)               & 2.3E-4 (1.3E-4$\endash$4.5E-4)   & 3.3E-3 (2.3E-3$\endash$6.8E-3)       & 3E3 ($500 \endash$9E3)   & $20$ ($15 \endash 24$)  & \editbfOne{$0.8$ ($0.1 \endash 3$)}\tablenotemark{\ensuremath{\star\star}}\\
CSS ($10^7 \le \Mstar \le 10^8\ \Msun$)               & 1.8E-3 (1.5E-3$\endash$2.9E-3)   & $0.020$ ($0.012 \endash 0.030$)      & $200$ ($60 \endash 600$)  & $30$ ($28 \endash 34$)  & \editbfOne{$0.15$ ($0.04 \endash 0.9$)}\tablenotemark{\ensuremath{\star\star}}\\
CSS ($10^8 \le \Mstar \le 10^9\ \Msun$)               & $0.024$ ($0.015 \endash 0.033$)  & $0.17$ ($0.099 \endash 0.27$)        & $2$ ($0.7 \endash 30$)    & $62$ ($42 \endash 76$)  & \editbfOne{$0.012$ (1.3E-3$\endash 0.5$)}\\
CSS ($\Mstar \ge 10^{9}\ \Msun$)                      & $0.41$ ($0.23 \endash 0.52$)     & $0.38$ ($0.31 \endash 0.49$)         & $3$ ($2 \endash 8$)       & $99$ ($82 \endash 110)$ & \editbfOne{$0.14$ ($0.03 \endash 0.4$)}\\
G1 (Mayall II)                                        & 4.6E-4    & 3.2E-3   & 7E3     & $26$  & $\editMathOne{5.1}$\tablenotemark{\ensuremath{\star\star}}\\
M32                                                   & $0.075$   & $0.11$   & $26$    & $76$  & $\editMathOne{0.48}$\\
M60-UCD1                                              & $0.018$   & $0.027$  & $450$   & $62$  & $\editMathOne{4.4}$\\
NGC 4486B                                             & $0.45$    & $0.18$   & $39$    & $170$ & $\editMathOne{8.1}$ \\
\hline
NGC 4342                                              & $3.3$     & $0.46$   & $17$    & $250$ & $\editMathOne{12}$\\
\hline
NGC 1277                                              & $12$      & $1.3$    & $3.5$   & $330$ & $\editMathOne{4.2}$
\enddata
\tablecomments{For E/S0 and CSS classes, the median value in the \citet{Norris14} sample is shown along with the 25--75\% quantile in parentheses. Note that the 25\%, 50\%, and 75\% quantiles pick different galaxies for different quantities in general. An ISO size distribution with $p = 2$ is assumed.}
\tablerefs{\citet{Trujillo14} (NGC 1277), \citet{Chabrier01,Anguiano18} (Local Galaxy), \citet{Norris14} (all others)}
\tablenotetext{\ensuremath{\star}}{$\sigmaOne = 30\ \kms$ is assumed, based on the local thin disk velocity ellipsoid \citep{Anguiano18}; it is not the central velocity dispersion of the Milky Way.}
\tablenotetext{\ensuremath{\star\star}}{\editbfOne{These rates may be understimated because gravitational focusing and acceleration by the planet and its host sun is neglected.}}
\end{deluxetable*}

The $\overline{\GammaImpact}$ for massive elliptical galaxies are of order $\editbfOne{0.04}\ \Gyr^{-1}$ (Table~\ref{table:MeanGamma}).  Similar results are obtained for the median $\Mstar \endash \sigma_e$ relation from the ATLAS-3D ETG sample, related to the Fundamental Plane \citep[][C13a]{Cappellari13-Size}.  There is a wide spread in $\overline{\GammaImpact}$, however.  Compact stellar systems (CSSs), including the most massive globular clusters, ultracompact dwarfs, and compact ellipticals \citep{Norris14}, generally have higher impact rates, driven mainly by stellar density rather than velocity dispersion.  \editbfOne{ISOs cause} $\overline{\GammaImpact} \approx 0.5\ \Gyr^{-1}$ in the Andromeda satellite M32, \editbfOne{which would imply a $1/4$ chance of an extra mass extinction} over the Phanerozoic.  In some extreme cases, $\la \editbfOne{100}\ \Myr$ are expected between impacts.  Finally, I considered NGC 1277 as a proxy for the ``red nugget'' population, a kind of compact massive ETG prevalent at high redshift \citep{Trujillo14}.  Because of its relatively high density and high velocity dispersion ($\sim 300\ \kms$), hazardous impacts are prevalent, \editbfOne{occuring about once every 200 Myr}.  The greatest $\overline{\GammaImpact}$ among the massive galaxies in the sample is the Virgo cluster galaxy NGC 4342, with properties intermediate between CSSs and red nuggets\editbfOne{, with one yottaJoule ISO impact per Earth-sized world every 80 Myr.}

If $\GammaImpact \ga 10\ \Gyr^{-1}$, then ISO impacts would likely dominate the mass extinction rate on planets with biosphere evolution similar to our own.  The time between impacts would also be of the same order as the time between the KT boundary and the evolution of humanity, and could inhibit ETI evolution if the appearance of humanity after Chicxulub is typical.  By that threshold, the majority of stars in most elliptical galaxies experience a ``safe'' flux of hazardous ISOs.  However, this is only a characteristic rate: stars near the center experience much higher rates (Section~\ref{sec:Models}).  The impact rate may be hazardous for large fractions of the stars in extreme CSSs and red nuggets.  This is interesting because \citet{Stojkovic19} proposed that high-metallicity dwarf galaxies including M32 are well-suited for ETI evolution.

\section{Impacts in the centers of large elliptical galaxies}
\label{sec:Models}

\subsection{Properties of Dehnen models}
Encounters are more common in the denser galactic cores. To model the radial dependence of $\GammaImpact$, I use D93 density profiles \editbfOne{at galactocentric distance $r$}:
\begin{equation}
\rho(r \equiv w a) = \frac{M}{a^3} \times \frac{(3 - \gamma)}{4 \pi} \frac{1}{w^{\gamma} (w + 1)^{4 - \gamma}} .
\end{equation}
The $\gamma$ index shapes the inner density profile, and $a$ is a scale radius marking the transition between the inner ($r^{-\gamma}$) and outer ($r^{-4}$) profile.  It is convenient to use the dimensionless radius $w \equiv r/a$, because the second multiplicand is independent of galaxy properties and only needs to be calculated once for each $\gamma$.  Although $\gamma$ is allowed to be in the range $0$ (fully cored) to $3$ (strongly cusped), I consider $\gamma = 1$ (Hernquist profile), $3/2$ ($R^{1/4}$ model), and $2$ (Jaffe profile).  Generally, giant boxy ellipticals with little net rotation have shallower (more Hernquist-like) inner profiles; disky ellipticals with fast rotation have steeper (more Jaffe-like) inner profiles \citep{Gebhardt96,Emsellem07}. As velocity dispersion in non-spherical profiles is not easily modeled, I assume spherical symmetry, with $r$ as the distance from the galaxy's center.  The mass $M$ can be either total mass, when calculating kinematic quantities, or stellar mass, when calculating properties of the stellar population.  \citet{Cappellari13-FP} find typical dark matter fractions within the half-light radius of $\sim 20\%$.  I ignore it as a relatively minor systematic compared to other uncertainties ($M = \Mstar$, $\rho = \rhoStar$).  The half-mass radius of the galaxy is located at $w_{1/2} = (2^{1/(3-\gamma)} - 1)^{-1}$ (D93).  

I \editbfOne{also} assume isotropic velocity dispersions for simplicity, although most elliptical galaxies have significant velocity anisotropy \citep{Cappellari07}.  D93 gives the 1D stellar velocity dispersion, assuming isotropy, as:
\begin{equation}
\sigmaOne^2 = \overline{v_r^2} = \frac{G M}{a} \times w^{\gamma} (w + 1)^{4 - \gamma} \int_{w}^{\infty} \frac{w^{\prime (1 - 2\gamma)} dw^{\prime}}{(1 + w^{\prime})^{7 - 2\gamma}} .
\end{equation}

The fundamental galaxy parameters, $M$ and $a$, are derived from the results of ATLAS3D \citep{Cappellari11}.  As a baseline, I apply the $\sigma_e$--$M_{\rm JAM}$ relation from C13a.  I also employ individual models of $258$ ETGs with measured stellar masses and velocity dispersions using the parameters in \citet{Cappellari13-FP}.  Since we observe elliptical galaxies in projection, neither the local density nor velocity dispersion is directly measured.  Instead, the projected observables are a surface brightness profile at projected distance $R$ from the center, $\Sigma(R) = \int_R^{\infty} \rhoStar(r) r dr / \sqrt{r^2 - R^2}$, and a line-of-sight integrated velocity distribution that is weighted by luminosity (and stellar mass): $\sigma_p^2 (R) = [2/\Sigma(R)] \times \int_R^{\infty} \rho(r) r \sigmaOne^2 dr / \sqrt{r^2 - R^2}$.  Furthermore, the velocity dispersion is averaged over some extended aperture due to limited angular resolution: $\sigma_e^2 (R) = [\int_0^R 2 \pi \Rprime \Sigma(\Rprime) \sigma_p^2 (\Rprime) d\Rprime]/[\int_0^R 2 \pi \Rprime \Sigma(\Rprime) d\Rprime]$.  The ratio of $\sigma_e^2$ and $G M / R_e$ is $0.15$ ($\gamma = 1$), $0.16$ ($\gamma = 3/2$), and $0.2$ ($\gamma = 2$), allowing me to relate \editbfOne{$R_e$ and $a$. The ATLAS3D data on $M$, $\sigma_e$, and $R_e$ implies a $\sigma_e^2 / (G M / R_e)$ consistent with $\gamma \approx 2$, but too high for $\gamma \approx 1.5$, especially for the largest slow rotator galaxies. Generally, using $M$ and $\sigma_e$ to derive $R_e$ results in relatively high $\rhoStar$, while using $M$ and $R_e$ results in relatively low $\rhoStar$. As a completely self-consistent Dehnen model requires me to choose one variable to derive, I calculate $M$ from $R_e$ and $\sigma_e$ because these are more directly observed parameters and the results are intermediate.}

\subsection{Results for models}

\begin{figure}
\centerline{\includegraphics[width=8.5cm]{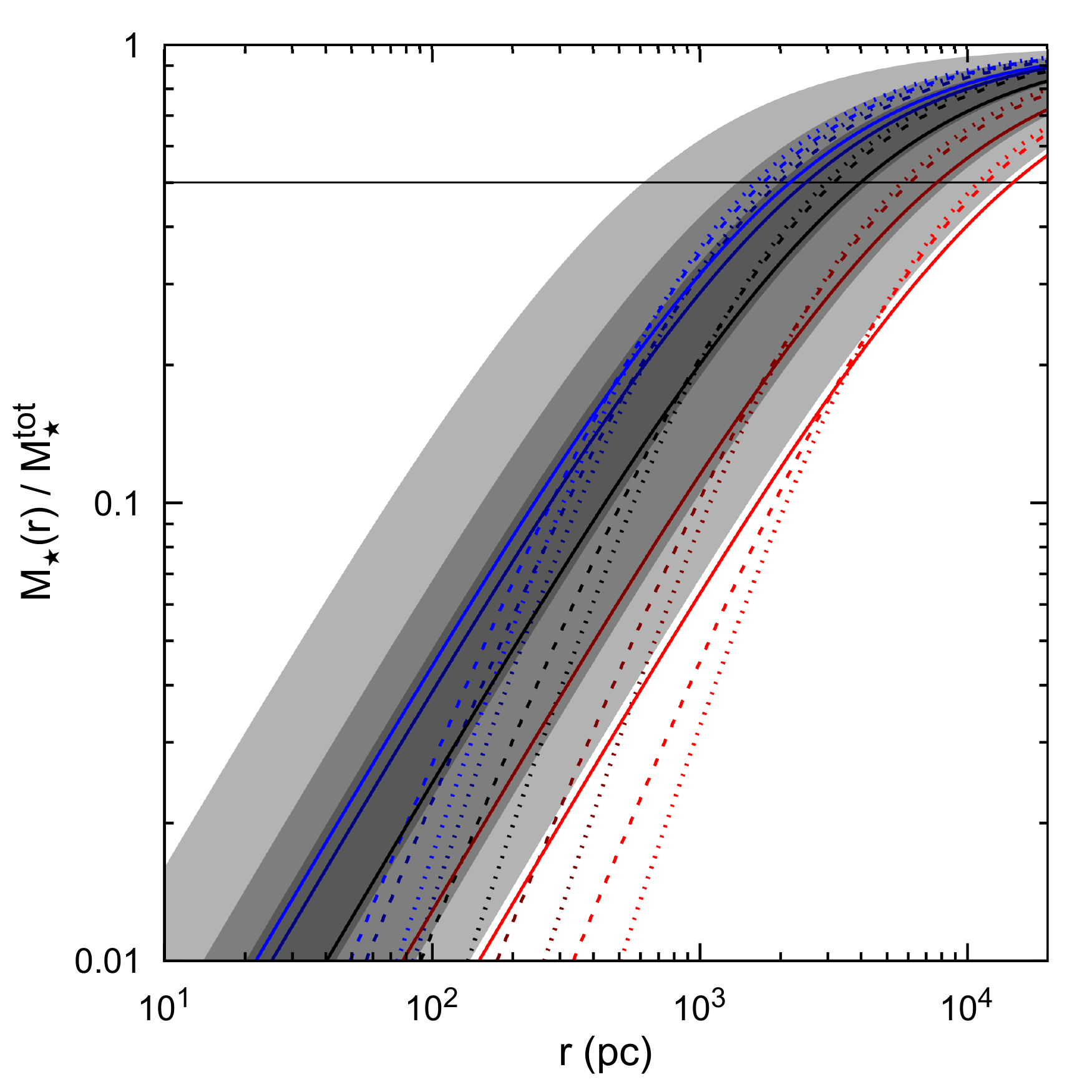}}
\figcaption{Estimated interior mass fraction as a function of radius in the Dehnen models. Models of galaxies on the C13a $\sigma_e \endash M_{\rm JAM}$ relation are plotted as line, with $\gamma = 1.0$ (dotted), $1.5$ (\editbfOne{dashes}), and $2.0$ (\editbfOne{solid}), and $\log (\Mstar/\Msun) = 10.0$ (bright blue), $10.5$ (dark blue), $11.0$ (black), $11.5$ (dark red), $12.0$ (bright red).  Shading shows the range for galaxies in the ATLAS3D sample (grey shading) using $\gamma = \editMathOne{2.0}$ models, with darker shading for those in the 5--95\% quantile, and darkest shading for the 25-75\% quantile. \label{fig:DehnenMFrac_C13}}
\end{figure}

The Dehnen models have power-law cusps in their centers. Unlike the exponential disk(s) of the Milky Way, this allows the density to rise arbitrarily high near the center, where a small but significant fraction of the stars in an elliptical galaxy may exist. The majority of stars are not in the deep interior (Figure~\ref{fig:DehnenMFrac_C13}): for $\gamma = \editMathOne{2}$, typical half-mass radii for the majority of the ATLAS3D sample are $\editMathOne{2.0} \endash \editMathOne{4.3}\ \kpc$, with very massive galaxies ($\Mstar > 10^{11}\ \Msun$) having larger radii. About 1--5\% of the stellar mass is within $100\ \pc$ of the center in most of the ATLAS3D galaxies.

\begin{figure*}
\centerline{\includegraphics[width=9cm]{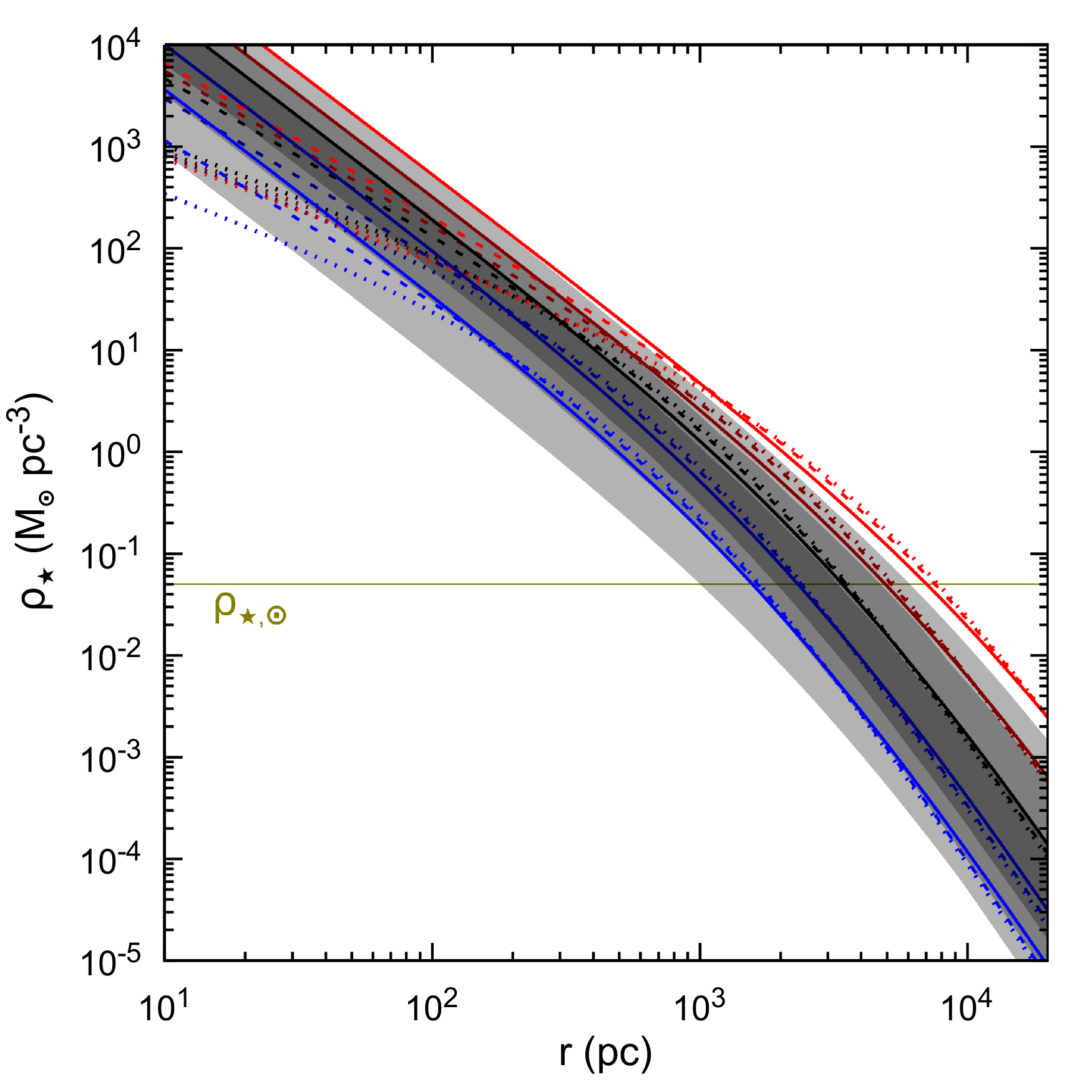}\includegraphics[width=9cm]{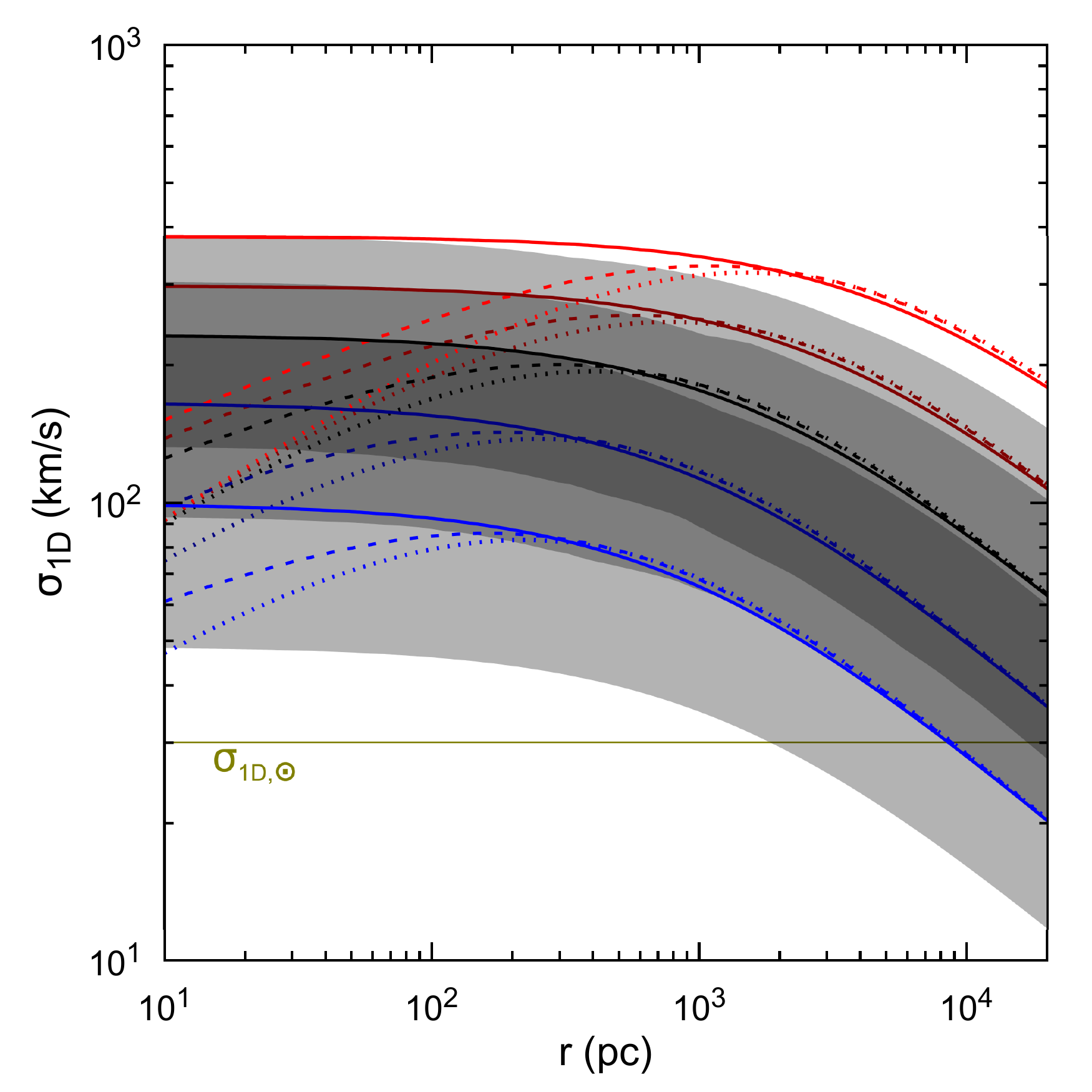}}
\figcaption{Estimated stellar mass density (left) and 1D isotropic velocity dispersion in the Dehnen models. Line styles and shading are the same as in Figure~\ref{fig:DehnenMFrac_C13}. The values for the Solar neighborhood are indicated by the gold lines.\label{fig:DehnenRho_C13}}
\end{figure*}

The typical stellar mass density at the half-mass radii is $\sim \editMathOne{0.01 \endash 0.1}\ \Msun\,\pc^{-3}$ (Figure~\ref{fig:DehnenRho_C13}), \editbfOne{of order that} in the Milky Way stellar disk.  However, a factor of $\sim 2 \endash 3$ decrease in $\editMathOne{r}$ results in the stellar density increasing by an order of magnitude; thus it typically reaches $\ga 100\ \Msun\,\pc^{-3}$ in the inner $100\ \pc$ in these models.  Stellar velocity dispersions at the half-mass radius are generally $\sim 100\ \kms$, reaching a peak a few hundred parsecs from the center when $\gamma < 2$ \editbfOne{and plateauing if $\gamma = 2$}.

\begin{figure*}
\centerline{\includegraphics[width=9cm]{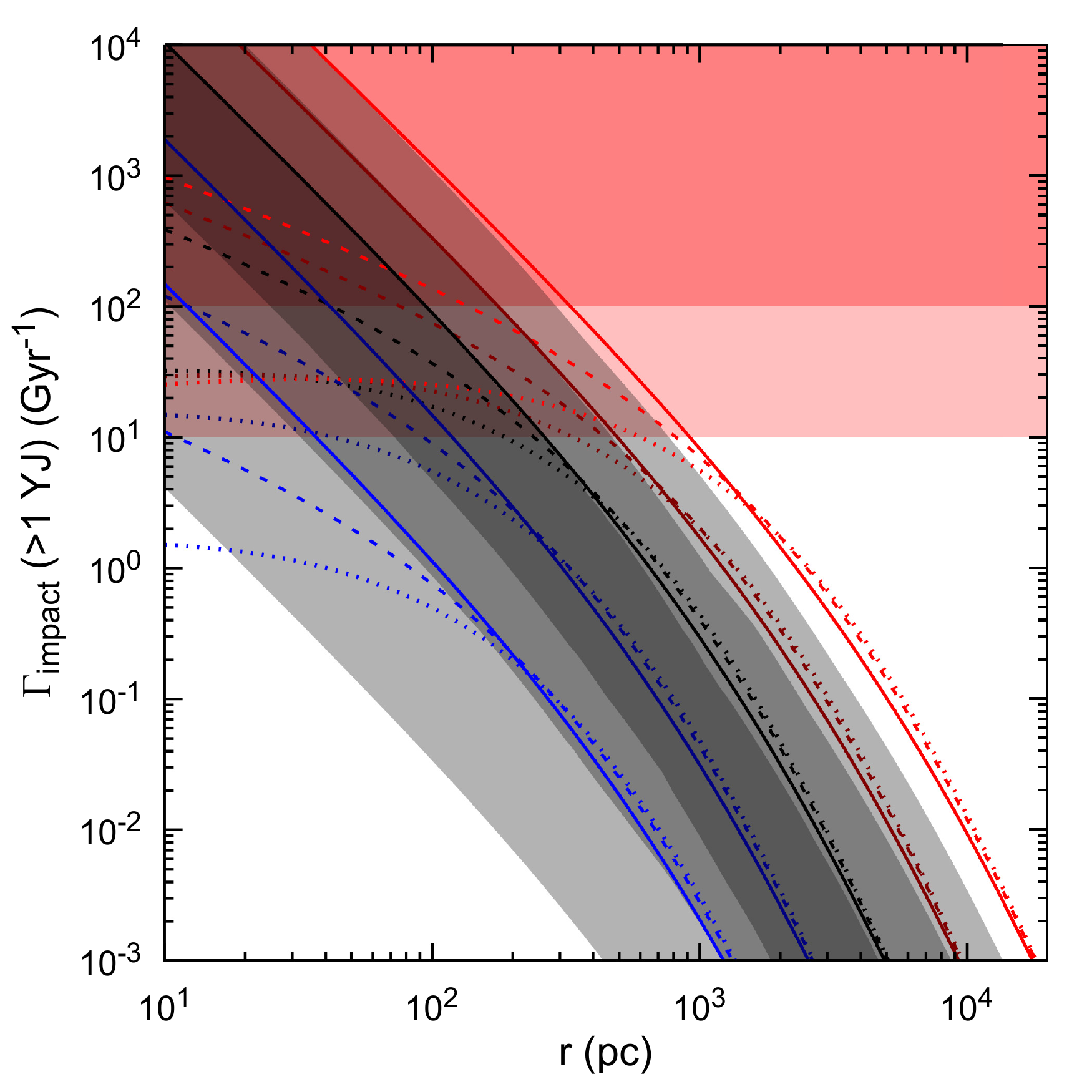}\includegraphics[width=9cm]{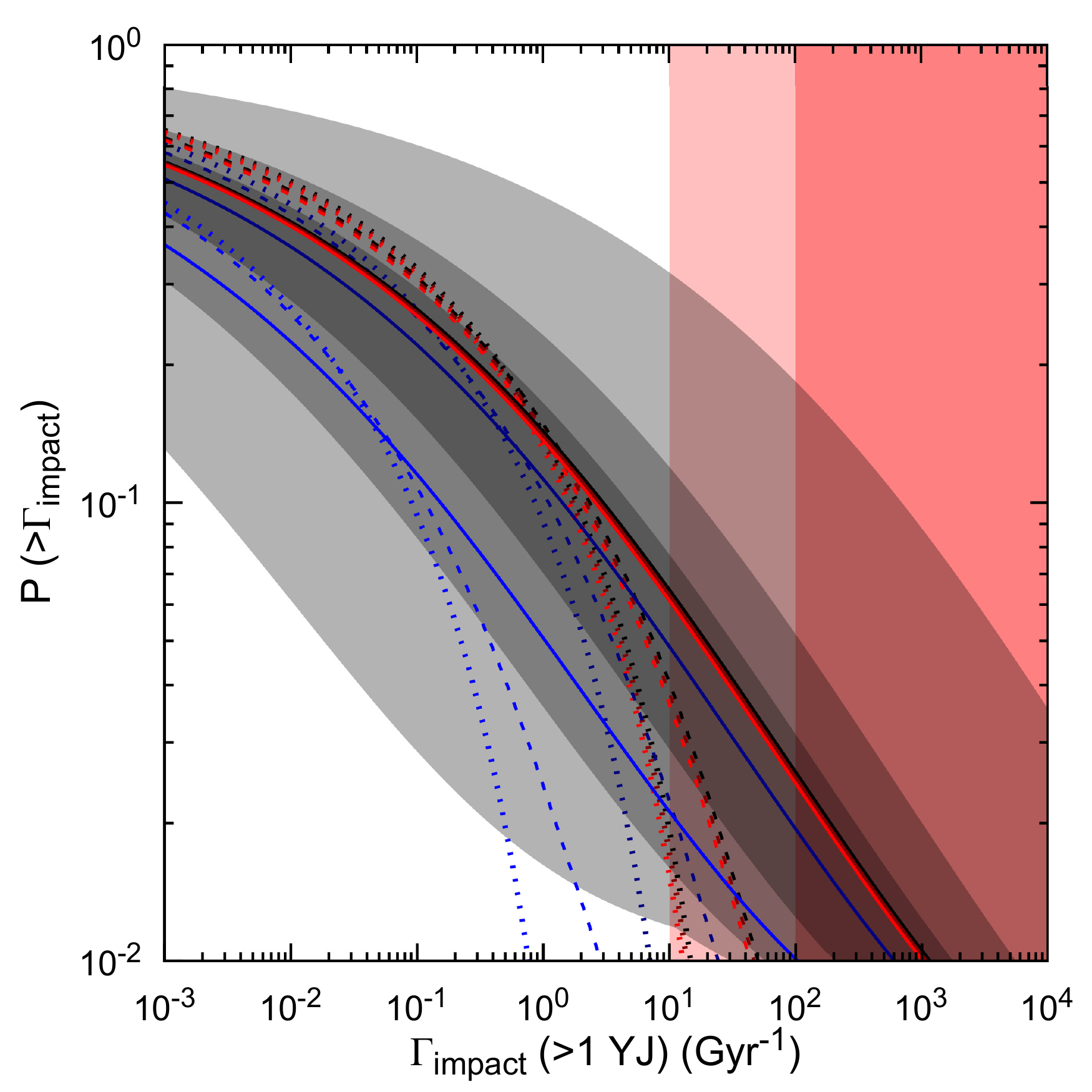}}
\figcaption{Estimated impact rates of hazardous ISOs in elliptical galaxies.    On left is the $\GammaImpact$ profile as a function of radius.  On right is the cumulative mass fraction of stars with $\GammaImpact$.  Red shading indicates hazardous impact rates, while grey shading and line styles are the same as in Figure~\ref{fig:DehnenMFrac_C13}. \label{fig:DehnenGamma_C13}}
\end{figure*}

I find that $\GammaImpact (\YJ)$ is $\la 0.01 \endash 0.1\ \Gyr^{-1}$ for the majority of stellar mass at any given moment (Figure~\ref{fig:DehnenGamma_C13}).  The impact rate increases rapidly towards the galactic centers, however, driven by the density cusp.  For massive ($\ga 10^{10.5}\ \Msun$) ellipticals lying on the C13a relation, a fraction $P_{10} \sim \editMathOne{5}\%$ of the stellar population closest to the galactic center experiences a YJ impact rate $\ge 10\ \Gyr^{-1}$.  Impact rates $\ga 100\ \Gyr^{-1}$ occur for the inner $P_{100} \sim \editMathOne{2}\%$ of the stellar mass in cuspy $\gamma = 2$ profiles.  Low mass elliptical galaxies have significantly reduced impact rates.  Similar conclusions are reached by considering the $258$ individual galaxy models, although with larger spread.  I find that half (90\%) of the galaxies have $P_{10} = \editMathOne{3 \endash 8} \%$ ($\editMathOne{1.6 \endash 12}\%$).

These impacts create exclusion zones for the evolution of highly complex (intelligent) life, albeit smaller than 1 kpc in radius.  Assuming a threshold of $\GammaImpact (\YJ) = 10\ \Gyr^{-1}$, the exclusion zone has a radius $\editMathOne{70 \endash 250}\ \pc$ ($\editMathOne{30 \endash 470}\ \pc$) in half (90\%) of the modeled ATLAS3D sample for $\gamma = \editMathOne{2}$.  Only a very small region, with radius $20 \endash \editMathOne{90}\ \pc$ \editbfOne{($10 \endash 180\ \pc$)} in half \editbfOne{(90\%)} of the $\gamma = \editMathOne{2}$ models, faces YJ impact rates $\ge 100\ \Gyr^{-1}$. \editbfOne{In $\gamma = 3/2$ models, the exclusion zone with $\GammaImpact (\YJ) \ge 10\ \Gyr^{-1}$ shrinks to $30 \endash 230\ \pc$ for half the modeled ATLAS3D sample.}

\begin{figure}
\centerline{\includegraphics[width=8.5cm]{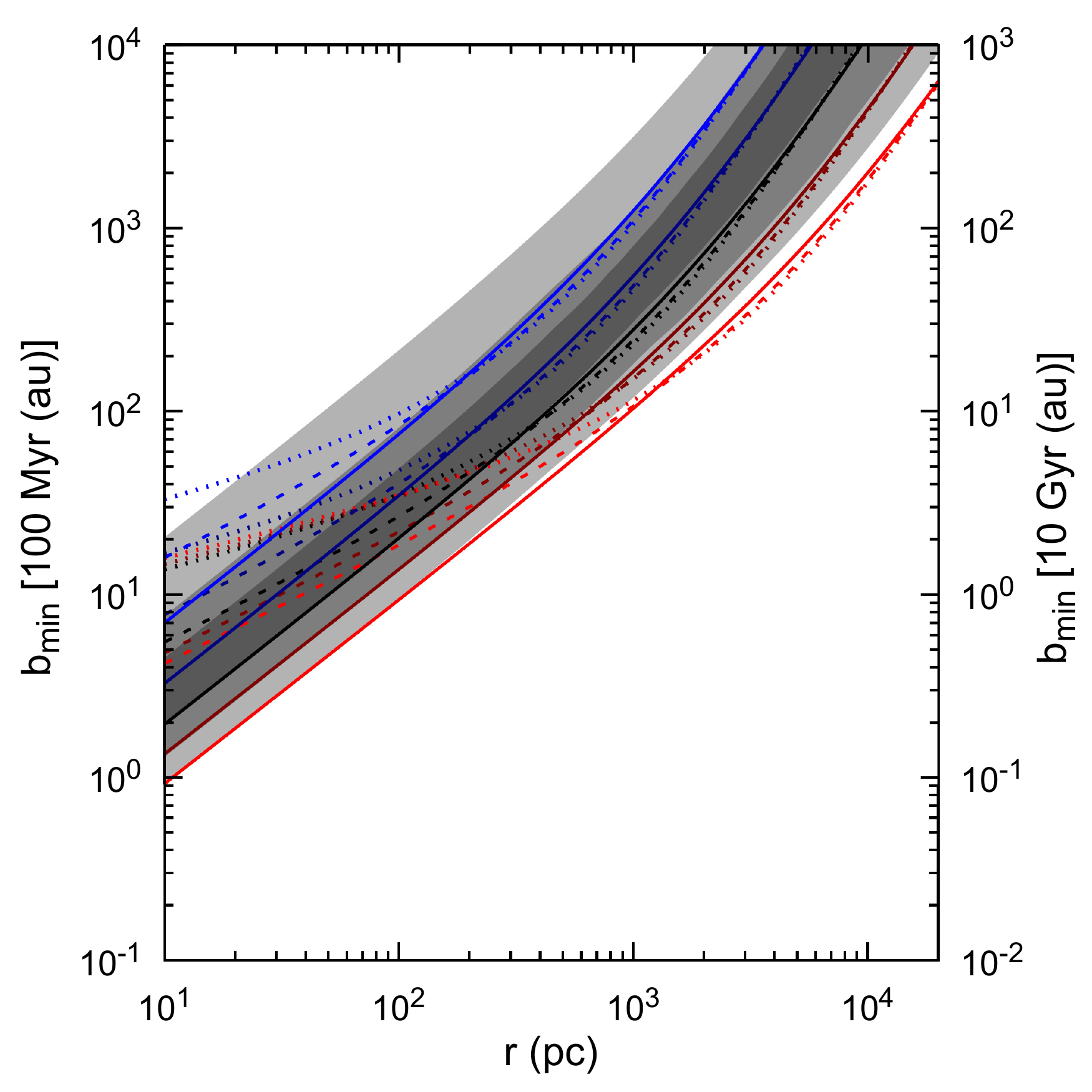}}
\figcaption{Estimated closest stellar passage $\bmin$ over 100 Myr (left) and 10 Gyr (right) as a function of radius in the Dehnen models.\label{fig:Dehnen_bmin_C13}}
\end{figure}

\editbfOne{Although it has been speculated that ejection of these objects into interstellar space is relatively inefficient \citep{MoroMartin09,Cook16}, minor bodies still bound to their host stars contribute to the flux of effective ISOs experienced by habitable planets. The impactors resulting from passage through debris belts can be included in $d\nISO/dm$.  If the radius $\adebris$ of these belts is smaller than the closest stellar passage expected over a duration $\GammaImpact^{-1}$, however, their impacts will generally occur in clusters separated by the typical time between passages through the debris belt $(\pi \nstar \adebris^2 \mean{\vrel})^{-1}$.  Figure~\ref{fig:Dehnen_bmin_C13} illustrates the expected closest stellar approach over 100 Myr in the Dehnen models. In half (90\%) of the $\gamma = 2$ models of the ATLAS3D galaxies, this distance is $34 \endash 56\ \AU$ ($25 \endash 72\ \AU$) at the edge of the nominal $\GammaImpact = 10\ \Gyr^{-1}$ zone, implying marginal clustering if most of the minor bodies are in Kuiper Belt-size disks. Further in, the closest approach distance decreases, and steady-state impactor flux is achieved for minor body populations closer to their stars.  At the edge of the $\GammaImpact = 100\ \Gyr^{-1}$ zone, the closest approach over 10 Myr is $11 \endash 19\ \AU$ ($8 \endash 25\ \AU$) in half (90\%) of these galaxies with $\gamma = 2$.}

These are only estimates, and there are several routes for further elaboration.  \editbfOne{Section~\ref{sec:OtherHab} considers habitability threats from the increased rate of stellar encounters, but o}ther potential dangers exist near galactic centers, including increased likelihood of being near a Type Ia supernova and proximity to nuclear activity.  More detailed dynamical models can examine the ISO flux variability that potential habitable planets face in their host stars' non-circular orbits.  Further models may also compare the relative abundance of planets and ISOs across the galaxy: planets may be biased towards the deadly but high metallicity cores, for example.  In addition, most elliptical galaxies are not spherically symmetric.  \editbfOne{Improved measurements of local ISOs and a better understanding of the minor body populations around other stars \citep[e.g.,][]{Trilling17,Seligman18,MoroMartin19-ExoOort} can help reduce the uncertainties in $\GammaImpact$, assuming that the populations in the Solar neighborhood are characteristic of those in elliptical galaxies.} Finally, a broader range of stellar systems may be considered, including CSSs and red nuggets (Section~\ref{sec:OoM})

\section{\editbfOne{Habitability threats from stellar encounters in elliptical galaxy cores?}}
\label{sec:OtherHab}
\editbfOne{The high ISO impact rates in the cores of elliptical galaxies are a combination of the high stellar densities ($\ga 30 \endash 100\ \Msun\,\pc^{-3}$ for $\gamma = 2$) and the high velocity dispersions ($\sim 130 \endash 200\ \kms$ for $\gamma = 2$) found in these regions. These same factors imply frequent close passages of stars through potentially inhabited stellar systems, with a typical closest approach over a time $t$ of $\bmin = (4 \sqrt{\pi} \nstar \sigmaOne t)^{-1/2}$:}
\begin{multline}
\label{eqn:bmin}
\editMathOne{\bmin = 3.1\ \AU\,\left(\frac{\nstar}{300\ \pc^{-3}}\right)^{-1/2} \left(\frac{\sigmaOne}{200\ \kms}\right)^{-1/2}} \\
\editMathOne{\times \left(\frac{t}{10\ \Gyr}\right)^{-1/2},}
\end{multline}
\editbfOne{where the given $\nstar$ corresponds to $\rhoStar \approx 100\ \Msun\,\pc^{-3}$ for a mean stellar mass $0.3\ \Msun$, as found in the inner $\sim 100 \endash 200\ \pc$ of large elliptical galaxies (see also Figure~\ref{fig:Dehnen_bmin_C13}). Stellar passages may directly affect a planet's habitability by knocking it out of the habitable zone \citep{DiStefano16,Kane18}, or indirectly endanger it by inducing orbital chaos by disturbing giant planets or triggering comet showers \citep{Hills81}.}

\editbfOne{Although the closest encounters are well within planetary systems, the high intruder speeds limit the time they spend there and the damage they can do \citep{Fregeau06}.  The encounter time $\bmin/\vrel$ of these closest passages are much shorter than the orbital period of an affected planetary body, and the passage is well within the impulsive regime of \citet{Spurzem09} (S09), with a typical eccentricity of $\sim 20$. Both the planetary body and its sun are briefly perturbed by the intruder star.  If the intruder passes one of them much closer than the other, then it is in the nontidal regime, with orbital changes resultant from the impulse on the body that the intruder star comes nearest to.  If the passage distance is greater than the distance $a$ between the planetary body and its sun, then it is in the tidal regime: because both the sun and the planetary body receive similar impulses from the intruder star, the relative velocity change between them is suppressed by a factor of order $\sim a/\bmin$ (S09). The frequency of dangerous encounters depends on the compactness of planetary systems, of course: the habitable zone of K dwarfs is significantly smaller, for example, requiring closer passages to pose a direct threat to the planets in it \citep[c.f.,][]{DiStefano16}. As a relatively conservative estimate, I will consider the effects on our Solar System.}

\begin{figure*}
\centerline{\includegraphics[width=9cm]{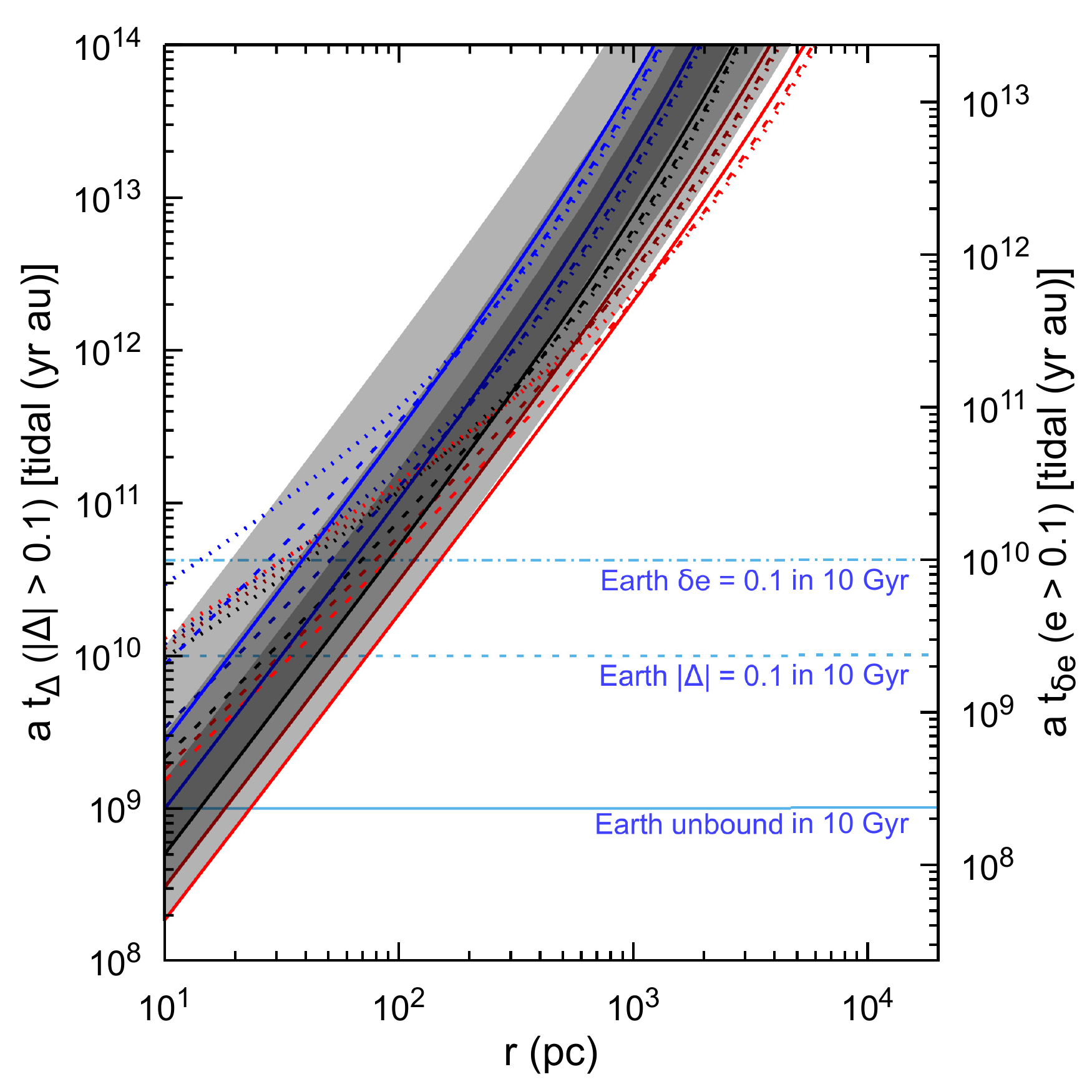}\includegraphics[width=9cm]{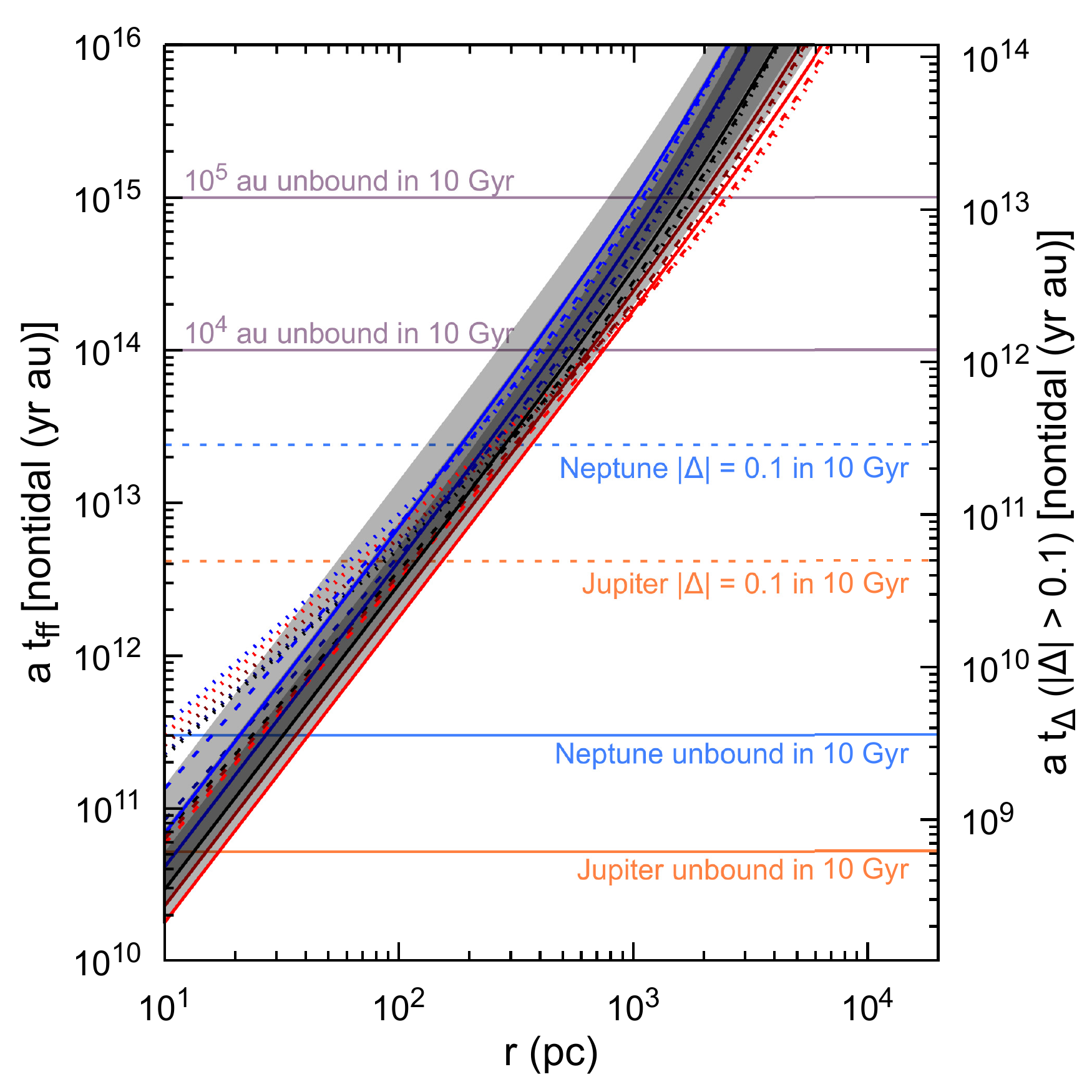}}
\figcaption{\editbfOne{Estimated time until a stellar passage perturbs a planet, scaled to a semi-major axis $1\ \AU$. On left, the tidal regime where $a \la \bmin$, which in a Solar System analog, would apply to Earth in most cases.  On right, the nontidal regime where $a \ga \bmin$.  In a Solar System analog, they would apply to the outer giant planets and minor bodies, as indicated by the lines indicating the time to perturb Jupiter, Neptune, or objects with Oort cloud-like $a$.} \label{fig:Dehnen_tFF}}
\end{figure*}

\editbfOne{Direct ejection of planets from the habitable zone should be very rare in elliptical galaxies outside nuclei (Figure~\ref{fig:Dehnen_tFF}). In the tidal regime characteristic of habitable planets around Sun-like stars, the cross-section for fractional changes of orbital energy $\Delta$ can be found by integrating equation 18 of S09. The time between encounters that induce $|\Delta| \ge \DeltaThresh$ is $\tDelta = 3 \DeltaThresh / (4 \pi \nstar \sqrt{G m_{\star} a^3})$, where the mass of the intruder star is assumed equal to the mass of the host sun, both with $m_{\star}$. S09 also yields an estimate for eccentricity variations. Starting from a circular orbit, a planet experiences a tidal encounter that raises its eccentricity by $\delta e$ about once every $\te = (\delta e)^2/(1.77 \nstar \sqrt{G m_{\star} a^3})$.  These timescales are $110\ \Gyr (\DeltaThresh/0.1)$ and $26\ \Gyr (\delta e/0.1)^2$ respectively for an Earthlike planet and $m_{\star} = 1\ \Msun$ in an $\nstar = 300\ \pc^{-3}$ environment with $\sigmaOne = 200\ \kms$. In over 90\% of the ATLAS3D galaxy models, $\te > 10\ \Gyr (\delta e / 0.1)^2$ at the edge of the $\GammaImpact = 10\ \Gyr^{-1}$ zone, with about a third satisfying this criterion at the $\GammaImpact = 100\ \Gyr^{-1}$ limit.  Figure~\ref{fig:Dehnen_tFF} shows these estimated times as a function of distance from the galactic center. In most elliptical galaxies, passages that induce $\delta e > 0.1$ in an Earth-analog are expected in 10 Gyr only within the inner hundred parsecs.  Unless the ecologies of habitable planets are extremely sensitive to their orbits, they are quite safe even throughout most of the cores of elliptical galaxies. Of course, the great majority of stars in large elliptical galaxies exist in much lower density regions, where the timescales are many trillions of years.  Even in large CSSs ($\Mstar \ga 10^7\ \Msun$) like M32, the timescales are generally tens of Gyr or longer, although as \citet{DiStefano16} notes, stellar encounters are likely a habitability threat in globular clusters.}

\editbfOne{In the inner hundred parsecs of large elliptical galaxies, the orbits of giant planets can be significantly perturbed over the history of a planetary system (Figure~\ref{fig:Dehnen_tFF}). The closest passages to these planets are in the nontidal regime. For $\DeltaThresh \ll 1$, the typical time between encounters inducing $|\Delta| \ge \DeltaThresh$ is $\tDelta \approx 3 \mean{\vrel} \DeltaThresh^2 / (32 \pi G a \rhoStar)$ (S09) or $2.1\ \Gyr (\DeltaThresh/0.1)^2 (a/(10\ \AU))^{-1}$ for the previously considered $\nstar = 300\ \pc^{-3}$ and $\sigmaOne = 200\ \kms$. These encounters are rarer than YJ ISO impacts on Earth-sized planets, although they may permanently endanger an inner planet's habitability. Furthermore, moderate changes in the giant planets' orbits may not sterilize a planet, as the giant planets in our Solar System may have migrated four billion years ago as in the Nice model without ejecting Earth from the habitable zone \citep{Tsiganis05}.}

\editbfOne{Frequent stellar encounters disrupt the structure of Kuiper Belt and Oort Cloud analogs in the cores of elliptical galaxies. The time between ``ionizing'' stellar encounters is $\tff = 3 \mean{\vrel} / (40 \pi G a \rhoStar)$ (S09). As shown in Figure~\ref{fig:Dehnen_tFF}, an comet cloud with $\mean{a} \sim 10^4\ \AU$ would be disrupted within 10 Gyr within the inner $\sim 500\ \pc$ of large elliptical galaxies.  During this unbinding process, a relatively small fraction of these comets may enter a loss cone trajectory that takes them into the inner solar system where they endanger habitable planets \citep{Hills81}. Kuiper Belt object analogs too may be perturbed into orbits that take them into the inner solar system in the hearts of elliptical galaxies. The relatively high number of these objects could make them even more dangerous than ISOs until they are depleted, although impact velocities will be much lower and the energy per impact much smaller.  A full accounting of this impact danger requires an understanding of the structure and diversity of debris belts in habitable planetary systems, and the Oort Cloud in particular remains poorly understood and largely unobservable even in our Solar System.}

\section{Conclusion}
Minor body impacts are a new component to our understanding of galactic habitability for complex life.  Because elliptical galaxies have high velocity dispersions, small ISOs can produce enormous devastation.  The impact rate should be particularly high near these galaxies' centers, where it can reach one every few Myr.  In elliptical galaxies with $\Mstar \ga 10^{10.5}\ \Msun$, I estimate \editbfOne{Earth-size worlds around} about $\sim \editMathOne{5}\%$ of the stellar mass at any given moment \editbfOne{are hit by} interstellar YJ impactors more than once every 100 Myr.  These would outpace Earth's Phanerozoic extinctions, possibly inhibiting the evolution of intelligent life.  Although big elliptical galaxies remain largely habitable, an exclusion zone occupies their inner few hundred parsecs.  Frequent impacts may occur in red nuggets and some CSSs. \editbfOne{Even throughout most of these dense zones, close passages of stars probably do not knock planets out of habitable zones and do not directly pose a habitability threat, although they may indirectly endanger these worlds by perturbing the orbits of outer giant planets and intrasystem comets.}

These impact provide a dramatic example of how galactic properties may directly control the evolution of biospheres.  Of course, they would not just happen on inhabited planets but all other worlds as well, shaping geological evolution too.  In the centers of these galaxies, all bodies experience intense cratering, with attendant heating and resurfacing.  

\acknowledgments
{\editbfOne{I thank the referees for their comments.} I \editbfOne{also} thank the Breakthrough Listen program for their support. Funding for \emph{Breakthrough Listen} research is sponsored by the Breakthrough Prize Foundation (\url{https://breakthroughprize.org/}).  In addition, I acknowledge the use of NASA's Astrophysics Data System and arXiv for this research.}

\bibliographystyle{aasjournal}
\bibliography{EllipticalComets_vProof_arXiv}

\end{document}